\documentstyle[aas2pp4,psfig2]{article}
\slugcomment{Submitted to ApJ}

\lefthead{J. J. Bryant & R. W. Hunstead}
\righthead{Host Galaxy ESO248-G10}

\begin{document}

\title{The Giant Radio Galaxy MRC B0319-454: Circumnuclear Structure of the 
Host Galaxy ESO248-G10}

\author{J. J. Bryant\altaffilmark{1}} 

\affil{School of Physics, The University of Sydney, NSW 2006, Australia.}

\and

\author{R. W. Hunstead\altaffilmark{2}}

\affil{School of Physics, The University of Sydney, NSW 2006, Australia}

\altaffiltext{1}{jbryant@physics.usyd.edu.au} 
\altaffiltext{2}{rwh@physics.usyd.edu.au} 

\begin{abstract}
We present optical and near-infrared images and spectra of ESO248-G10,
the host of the giant radio galaxy MRC B0319-454. From near-infrared
colours, the active nucleus is shown to be reddened by hot dust emission 
or dust extinction. Star forming regions
are identified beyond a radius of 5\,arcsecs (8.1\,kpc) where hot gas
is prevalent. The
optical spectrum shows gas rotation at speeds of up to $\sim350$\,km\,s$^{-1}$
out to $\geq32$\,kpc along the radio axis.
A model is proposed in which the
giant elliptical is triaxial with the radio axis along the short
axis, and the figure rotation is around the long axis. From the model,
the angles of the principal axis are 
$\psi=34^{\circ}$, $\theta=65^{\circ}$, and $\phi=19^{\circ}$ with axis
ratios of $B/A=0.75$ and $C/A=0.69$.
A second velocity feature from 5 to 15\,arcsecs to the north-east of
the nucleus is proposed to be a merging gas-rich galaxy inducing star
formation while settling into an orbit about the figure rotation axis.

\end{abstract}


\section{INTRODUCTION}
\label{intro}

Radio galaxies are frequently found to have giant ellipticals as
their hosts (Carballo et al. 1998 and references therein).
In elliptical hosts with a dust lane, the radio 
jets are typically aligned perpendicular to the dust lane
(Kotanyi \& Ekers 1979; Mollenhoff, Hummel \& Bender 1992).
The sense of rotation 
of the dust lane about the radio jet axis can be parallel, 
anti-parallel or perpendicular to the stellar rotation in a stable system.
This implies an external origin for the dust.
The common occurrence of dust lanes in elliptical galaxies suggests that
infalling dust has been trapped in stable closed orbits.
It is not surprising, therefore, that Heckman et al. (1986) found that 50$-$75\% of radio 
galaxies show evidence of ongoing or past mergers$/$interactions.
The orientation of the radio axis and rotation of a dust lane compared
to the stellar rotation are indicative of the merger history of the
galaxy.

Many elliptical galaxies show evidence of complex kinematics. Warped 
dust lanes, radio jets misaligned with the major
and minor isophotal axes, and 
tumbling around an axis perpendicular to the radio axis are all indicative
of triaxiality in elliptical galaxies. While a warped dust lane could indicate
that the dust has not yet settled into a preferred plane, van Albada,
Kotanyi \& Schwarzschild (1982) have modelled warped dust lanes as stable orbits
in tumbling triaxial systems.
Binney (1978) first proposed that elliptical galaxies might be triaxial.
Since then it has come to be accepted that most, if not all, ellipticals
are triaxial and many also show evidence of tumbling.
Unfortunately, the intrinsic shape and orientation of such a system cannot
be determined from surface photometry alone. While the apparent axis ratios
can put limits on the lengths of the intrinsic axes,
stellar and ionised gas kinematics together with the dust lane orientation are 
essential for the determination of a triaxial shape, as they
trace the gravitational potential.

MRC B0319$-$454 is a giant radio galaxy with an FR II (Fanaroff \& Riley 1974) morphology. It was 
identified by 
Jones (1989) with the Molonglo Observatory Synthesis Telescope (MOST).
The double-lobed radio structure has a projected separation of 2.51\,Mpc 
(H$_{0}=50$\,km\,s$^{-1}$\,Mpc$^{-1}$), with quasi-continuous
jets at a position angle of $51^{\circ}$ and radio knots mapped by Saripalli, Subrahmanyan \& Hunstead (1994). 
The host galaxy, ESO248-G10, has been identified with a 
giant elliptical galaxy at a redshift of 0.0625, giving a
distance of 337\,Mpc. At this distance 1\,arcsec corresponds to 1.63\,kpc.
With a major axis of $\sim100$\,kpc at the $R=24.5$\,mag\,arcsec$^{-2}$ 
contour, it is an ideal
candidate to investigate the host galaxy properties of a giant radio galaxy.

The optical images of Saripalli et al. (1994) show ESO248-G10 to have a warped dust lane.
The radio axis is misaligned with the optical isophotal minor
and major axes by $34^{\circ}$ and $56^{\circ}$ respectively, hinting that the galaxy
may be triaxial. In this paper we attempt to deduce the true orientation 
of the principal axes to the apparent isophotal axes from the dust lane
morphology and ionised gas dynamics.
Furthermore, colour analysis is used
to find direct evidence of a merger or interaction.
Section~\ref{Obs} describes the observations and reduction, with the results
in section~\ref{Results}. A discussion of the dynamics and
orientation of ESO248-G10 is in section~\ref{Discussion}.

\newcommand{\blank}{\vspace{\baselineskip}}

\section{OBSERVATIONS AND REDUCTION}
\label{Obs}

\subsection{Images}

The near-infrared bands of {\it J}, {\it H}, and {\it Kn} are ideal 
for studying the circumnuclear regions of ESO248-G10 as infrared wavelengths
can penetrate the enshrouding dust.
Mosaiced images centred on the nucleus of ESO248-G10 were taken in {\it J}$-$, {\it H}$-$,
and {\it Kn}$-$bands (see Table~\ref{TObs}). The {\it Kn} image was obtained with IRIS (Allen 1992) at the
f$/15$ cassegrain focus of the 3.9\,m Anglo-Australian Telescope (AAT) at Siding Spring
Observatory on 1994 October 22. Pixels were 0.6 arcsec on the sky, with
$128\times128$ pixels on the HgCdTe array. The mosaic was comprised of
three, five-point patterns centred on the {\it Kn} point source with offsets of
between 10 and 15 arcsec for the surrounding frames. 
The standard star HD20223 was observed
immediately after the image frames. Sky frames were offset by between 80
and 190 arcsec from the galaxy centre.

\begin{table*}
\centering
\begin{minipage}{140mm}
\caption{Log of Observations.}
\begin{tabular}{@{}lccccc}
\tableline
\tableline
 & & &Av. Seeing & & Integration Time \\
 UT Date    & Type  &$\lambda$ & (arcsec) &Photometric &  (s) \\
\tableline
1994 Oct 22  &image  & Kn & 1.4 & Yes & 750 \\
1999 Feb 4  &image  & H  & 1.3 & Yes & 1000 \\
1998 Oct 1-2  &image  & J  & 1.5 & Yes & 1740 \\
1995 Feb 7  &spectrum  & Opt.  &1.25 & Yes & 2000 \\
1995 Oct 31 &spectrum & HK & 2.5 & No & 11640 \\
\tableline
\end{tabular}
\label{TObs}
\end{minipage}
\end{table*}

{\it J} and {\it H} images were taken with CASPIR (McGregor et al. 1994) on the MSSSO 2.3\,m
telescope at Siding Spring Observatory on 1998 October 1-2 and 1999 February 4 
respectively. The $256\times256$ InSb array
was used in FAST mode with 0.5\,arcsec pixels.
A nine-point pattern of images made up both the mosaics, with offsets of
8\,arcsec for each frame surrounding the nucleus. 
Sky frames were offset by 300\,arcsec from the nucleus. The standard stars
HD20223 and HD15911 were observed immediately before or after the image
frames.

Data reduction and image processing was done with four software packages: {\sc iraf} 
(Tody 1993),
{\sc figaro} (Shortridge 1993), {\sc karma} (Gooch 1996), and {\sc miriad} 
(Sault, Teuben \& Wright  1995).
The {\it Kn}$-$band IRIS image was reduced with the IRIS-specific tasks in 
{\sc figaro}, while the {\it J} and {\it H} images used the CASPIR reduction
package in {\sc iraf}. In each case the frames were flat fielded using
dome flats and then sky-subtracted. All three wavebands were cleaned of bad
pixels with
{\sc figaro clean} and mosaiced with {\sc irismos}. At least four stars in the
field were used to register the frames when mosaicing the {\it J} and {\it Kn} images.
The {\it H} image, however, was taken with tip-tilt correction and 
blind mosaicing was sufficient. All three images were taken
in photometric conditions and were magnitude calibrated using the standard
star observations.

\subsubsection{Alignment of Images}

The {\it J}, {\it H}, and {\it Kn} images were aligned relative to each other by the 
following method. The {\it Kn} image was resampled from 
0.6\,arcsec\,pixel$^{-1}$ to 0.5\,arcsec\,pixel$^{-1}$ with {\sc iraf magnify} to match 
the {\it J} and {\it H} images.
By cross-referencing six stars on the digitized sky survey (DSS\,I, blue) image in {\sc karma koord}, 
coordinates were attached to the {\it J} mosaic accurate to 0.1\,pixel rms.
Coordinates were transfered to the {\it H} and {\it Kn} images in the same 
way but the {\it Kn} image was smaller so that fewer alignment stars gave a 
larger scatter. It was then necessary
to shift the {\it H} and {\it Kn} images with {\sc iraf imshift} to align 
the relative pixels to the {\it J} image to $<0.2$\,pixel.
$J-H$ and $H-Kn$ magnitude images were then made in {\sc miriad} by subsetting
an identical region from each of the images, then converting to magnitude
images and subtracting.
A coordinate system was applied to $H-Kn$ and $J-H$ images by copying the 
fits header parameters from the parent images.                                             

The {\it B} and {\it R} CCD images were those used by Saripalli et al. (1994).
They were taken at the AAT, and are measured to have a seeing disk FWHM of 
2.65\,arcsec and
1.8\,arcsec for {\it B} and {\it R} respectively. 
Coordinates were transferred to both images by comparison with the DSS
image. The resulting images align with the {\it J} image to $<0.2$\,arcsec.
The {\it R} image was convolved to have to same resolution as the {\it B} 
image, before a $B-R$ image was made with the same relative astrometric accuracy.

\subsection{Spectra}
\subsubsection{Optical}
\label{spectraobs}
\begin{figure*}
\centerline{\psfig{figure=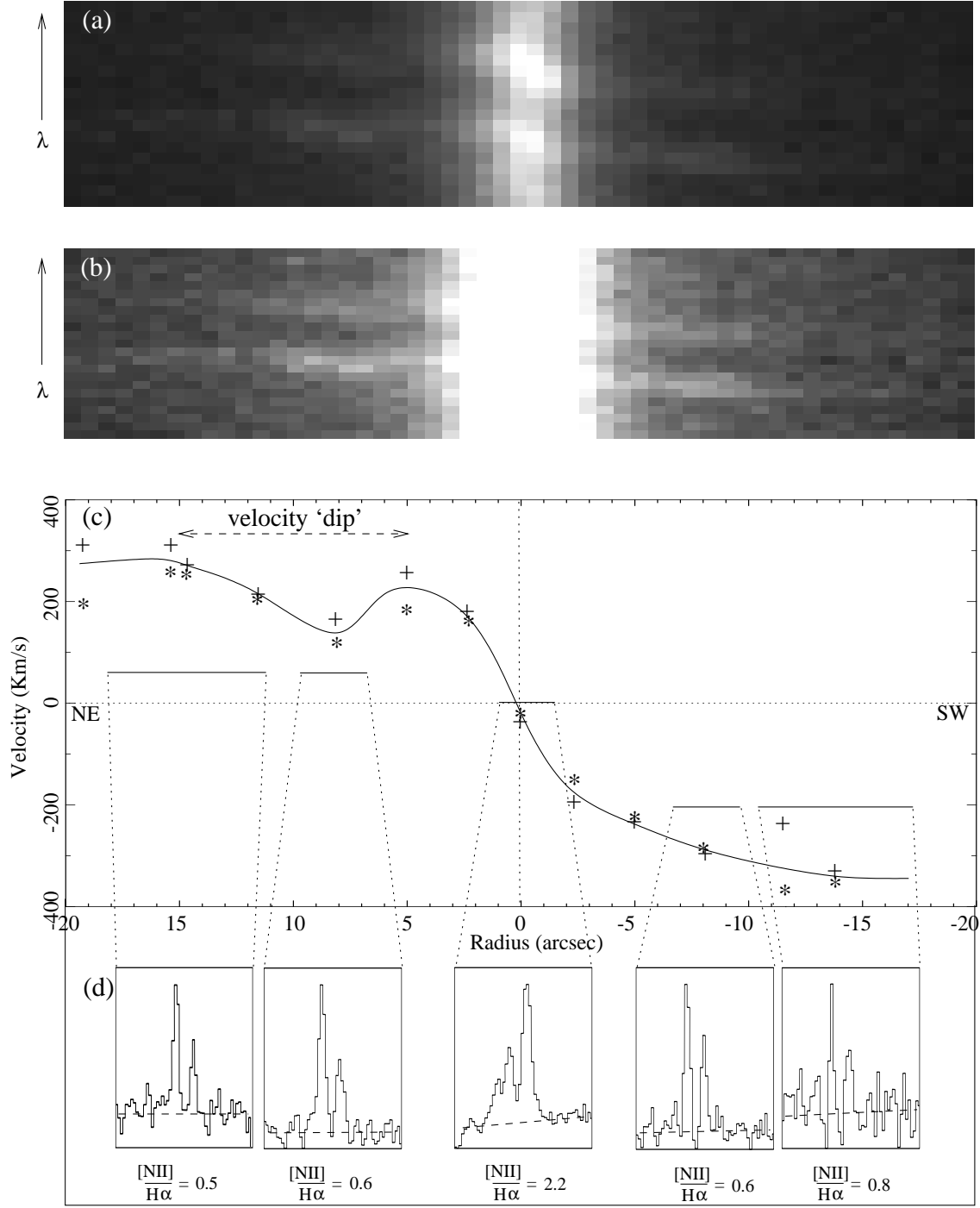,height=20cm}}
\caption{(a)Optical Spectral Image including the
[N\,{\sc ii}]$\lambda6584$ and H$\alpha$ 
lines. The inner pixels clearly show fast rotation of $\sim100$\,km\,s$^{-1}$\,arcsec$^{-1}$.
(b) Same as (a) with greyscales chosen to highlight the extent of the
rotating gas out to almost 20\,arsecs.
(c) Rotation Curves for [N\,{\sc ii}] (crosses) and H$\alpha$ (stars) 
along p.a. of $40^{\circ}$. 
See text for explanation of the points.
(d) Change in the relative strength of 
[N\,{\sc ii}]$\lambda6584$ and H$\alpha$ 
with position. Each spectrum is for the region marked by the
corresponding line on the rotation curve and has a width of $\sim182$\,\AA.}
\label{rotcurve}
\end{figure*}

Optical spectra were taken for us as service observations at the AAT on
1995 February 7 (see Table~\ref{TObs}).
The RGO spectrograph was used with the 25cm camera, 270R grating, and 
Tektronix 1k\,$\times$\,1k CCD.
The slit was 192.5\,arcsec long with a spatial scale of 0.76\,arcsec\,pixel$^{-1}$;
it was aligned at
a position angle of $40^{\circ}$, rather than the requested
$51^{\circ}$ (radio jet axis p.a.)
and consequently is not along
any physically significant direction. Sky subtraction and cosmic ray removal
was done in {\sc figaro}. 
Wavelength calibration was done with CuAr lamp spectra and the measured spectral resolution is 7.5\AA\, FWHM.
The resultant spectral image shows significant
rotation in the stronger emission lines 
[N\,{\sc ii}]$\lambda6584$ and H$\alpha$, which can be seen
in Figure~\ref{rotcurve}(a)\&(b). Spectra were therefore extracted 
along the slit
to measure the velocities either side of the nucleus.
To improve the signal-to-noise ratio away from the nucleus an increasingly wide
extraction window was used.
A 3-pixel spectrum was extracted at the centre,
then spectra with 3, 4, 4, 5, 5, and 5 pixels were extracted consecutively
to the north-east, and 3, 4, 4, and 5 pixels similarly to the south-west.
Two more spectra of 9 pixels
each were centred 19 pixels north-east and 18 pixels south-west of the centre.
Each of the resulting spectra was divided by that of
the standard star HD26169.
From each of these spectra, the [N\,{\sc ii}]$\lambda6584$
and H$\alpha$ lines were deblended with {\sc iraf ngaussfit} by constraining the
relative flux, position and width of the [N\,{\sc ii}]$\lambda6584$
\& 6548 lines. Other lines were measured with {\sc iraf splot}.
Figure~\ref{rotcurve}(c) shows the rotation curve formed from 
the [N\,{\sc ii}]$\lambda6584$ and H$\alpha$ line velocities measured
from each of the extracted spectra. Each point on the curve is plotted at the center of the
extracted window that formed each of the 13 spectra.
It is noteworthy that the lines could be measured to a radius of
$\sim20$\,arcsec
to the north-east, but only $\sim14$ arcsec to the south-west.
The S$/$N in the MgIb line was not sufficient to measure a stellar rotation
curve.

Optical extinction is typically measured from the H$\alpha$/H$\beta$
ratio. The H$\beta$ line is, however, not measurable with sufficient accuracy
to give a meaningful estimate of extinction. Furthermore, as the nucleus is below the highest
obscuration of the dust lane (see section~\ref{ResultsImages}), extinction correction was not applied to
the emission line measurements. The errors in the 
[N\,{\sc ii}]$\lambda6584/\rm{H}\alpha$ line ratios due to 
extinction are negligible as the wavelengths are close.
Figure~\ref{rotcurve}(d) shows the 
[N\,{\sc ii}]$\lambda6584/\rm{H}\alpha$ 
ratios for five of the spectra. These ratios will be discussed further
in section~\ref{AGN}.

\subsubsection{Near-infrared}

An HK echelle spectrum was obtained with IRIS on the AAT on 1995 October 31.
The wavelength coverage is from 1.45 to 2.5\,$\mu$m, with a resolution of
400. The slit was 13 arcsec long, with a spatial scale of 0.79 arcsec
pixel$^{-1}$, and oriented east-west.
The galaxy nucleus was positioned 3.5 arcsec either side of the centre
of the slit in alternate frames. Typical IRIS reduction involves subtraction
of pairs of frames to remove sky and background galaxy. While this method
was used here, it was discovered later (see section~\ref{AGN}) that beyond
4\,arcsec from the nucleus along the optical slit angle, the galaxy has
starburst processes not found at the nucleus. Starbursts regions may also be 
present in the direction of the HK slit, at a radius of 7 arcsec from the 
nucleus, in the region used for sky subtraction.
There is therefore a chance that the sky background
spectra that were subtracted may have starburst lines not found at the nucleus. However,
contribution from these possible lines is expected to be small, as the nucleus is so 
much brighter than the background galaxy. 
This possible contamination is taken into account in evaluation of the spectrum.

The echelle frames were flat-fielded and cleaned in {\sc figaro}. After
coadding, spectra were extracted from the central 4 pixels. Two standard stars
bs1006 and HD21473 were used to flux calibrate the spectrum with {\sc irisflux}. As
the observing conditions were not all photometric, this flux calibration is only
relative. Wavelength calibration was done with argon lamp spectra and 
wavelengths were corrected to a heliocentric frame.

\section{RESULTS}
\label{Results}

\subsection{Images}
\label{ResultsImages}

\begin{figure*}
\centerline{\psfig{figure=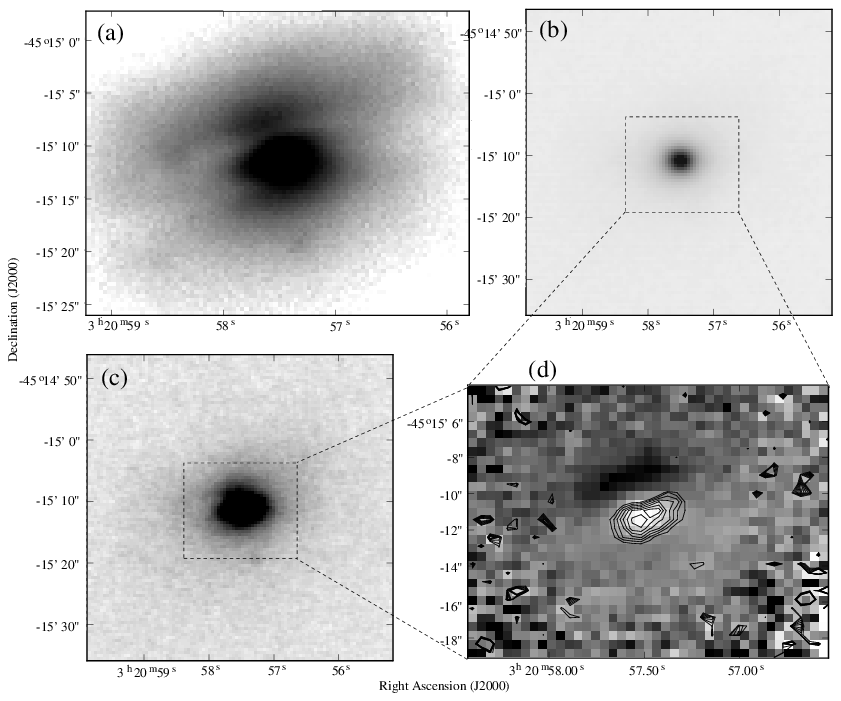,height=16cm}}
\caption{(a)$R$, (b)$Kn$, and (c)$J$ band images. (d) $H-Kn$ extinction image,
with overlaid contours of the $J-H$ extinction image; darker regions indicate
greater reddening. Contour levels are 
0.86-0.98 in 0.02 steps.  }
\label{images}
\end{figure*}

In Figure~\ref{images}, the {\it R} image shows a dust lane crossing the
front of the galaxy to the north-east of the nucleus. This suggests the
galaxy is tilted away from the sky plane to the north-east.
The nucleus is therefore
positioned below the strongest band of extinction.
The near infra-red images are much more centrally condensed, with the
{\it Kn}$-$band image having an unresolved core coincident, within errors,
with the radio core reported by Saripalli et al. (1994).
The {\it J} image shows a notch-shaped indentation approximately 4\,arcsec 
to the east of the nucleus, which is not
apparent in the {\it H} and {\it Kn} images. Overlaid $J-H$ and $H-Kn$ extinction images formed from the
division of the $J$, $H$, and $Kn$ images are shown in Figure~\ref{images}(d).
As expected, the strongest reddening
in the $H-Kn$ image is above the nucleus, along the dust lane. 

Aperture photometry results are given in Table~\ref{apphot}. Apertures were
chosen to extend well past the north-east edge of the dust lane, with measurements
made using the {\sc iraf radprof} program. All apertures are centred on the
$Kn-$band peak, which coincides, within errors, with the peaks in $J-$
and $H-$bands. As a background aperture is required for photometry, and
the three images are different sizes, an annulus was chosen that fits
within each image and avoids any stars. With inner and outer radii
of 36 and 41\,arcsec, this
background aperture avoids the main galaxy halo, but a small contribution
from the galaxy cannot be ruled out. At large apertures, this introduces
an estimated systematic error of up to 0.1\,magnitudes. In smaller apertures, however,
the error is dominated by standard star photometry errors, which are 
0.02\,magnitudes.

\begin{table*}
\centering
\begin{minipage}{90mm}
\caption{Aperture photometry centred on the IR point source. Errors 
in the colours range from $0.02-0.1$\,mag from smallest to largest apertures.}
\begin{tabular}{@{}lccccc}
\tableline
\tableline
Aperture & & & & & \\
Radius & {\it J}  & {\it H} & {\it Kn} & $J-H$ & $H-Kn$ \\
(arcsec) &(mag)  &(mag) &(mag) &(mag) &(mag) \\
\tableline
 1 & 15.24 &14.30 &12.66 &0.94 & 1.65\\
 2 & 14.24 &13.35 &11.47 &0.89 & 1.89 \\
 3 & 13.82 &12.96 &10.94 &0.86 & 2.02 \\
 4 & 13.54 &12.71 &10.64 &0.83 & 2.07 \\
 5 & 13.35 &12.54 &10.45 &0.81 & 2.09 \\
 6 & 13.19 &12.40 &10.30 &0.79 & 2.10 \\
 7 & 13.07 &12.30 &10.18 & 0.77 & 2.11 \\
 8 & 12.96 &12.21 &10.09 & 0.76 & 2.12 \\
 9 & 12.88 &12.13 &10.01 & 0.74 & 2.13 \\
10 & 12.80 &12.07 & 9.94 & 0.73 & 2.13 \\
\tableline
\end{tabular}
\label{apphot}
\end{minipage}
\end{table*}

Aperture photometry in {\it Kn}-band gave the absolute magnitude 
in a 1.5\,arcsec radius aperture as M$_{K}=-25.9\pm0.5$.
The canonical luminosity criteria for optical QSOs is 
M$_{K}=-25.8$ (Surace \& Sanders 1999),
while typical Seyferts only have M$_{B}=-20.2$.
Therefore, the infrared light from ESO248-G10 is sufficient to class it
with the QSOs. 

\subsection{Spectra and Rotation Curves}
\subsubsection{Emission Lines and Ratios}
\label{EmissionL}
\begin{figure*}
\centerline{\psfig{figure=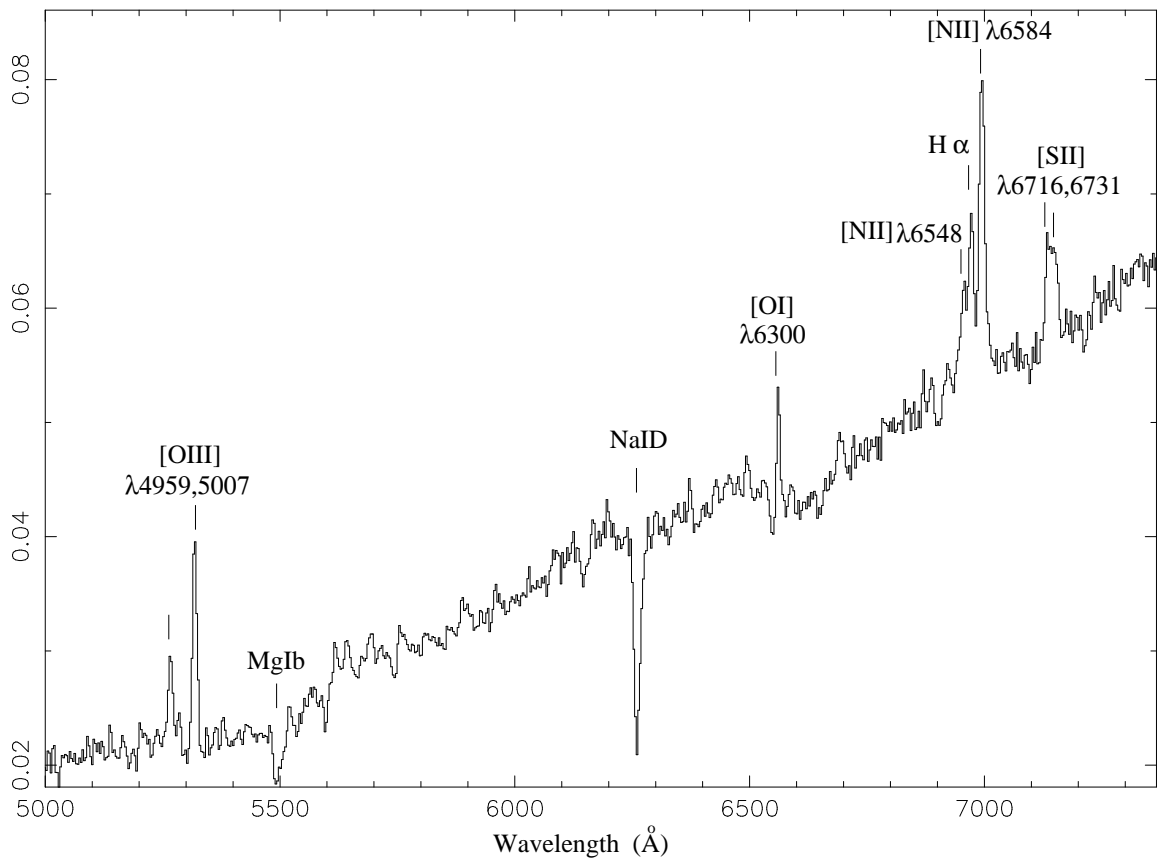,height=13cm}}
\caption{The optical spectrum of ESO248-G10 at the nucleus. Relative
values only are shown on the intensity scale as the spectrum is not flux
calibrated.}
\label{spectrum}
\end{figure*}

Figure~\ref{spectrum} shows the optical spectrum at the nucleus.
The emission lines of 
[N\,{\sc ii}]$\lambda\lambda6548, 6584$, H$\alpha$, 
[O\,{\sc iii}]$\lambda\lambda4959, 5007$, [O\,{\sc i}]$\lambda6300$, 
and [S\,{\sc ii}]$\lambda\lambda6716, 6731$,
and the absorption lines of NaID and MgIb are all marked. 

\begin{figure*}
\centerline{\psfig{figure=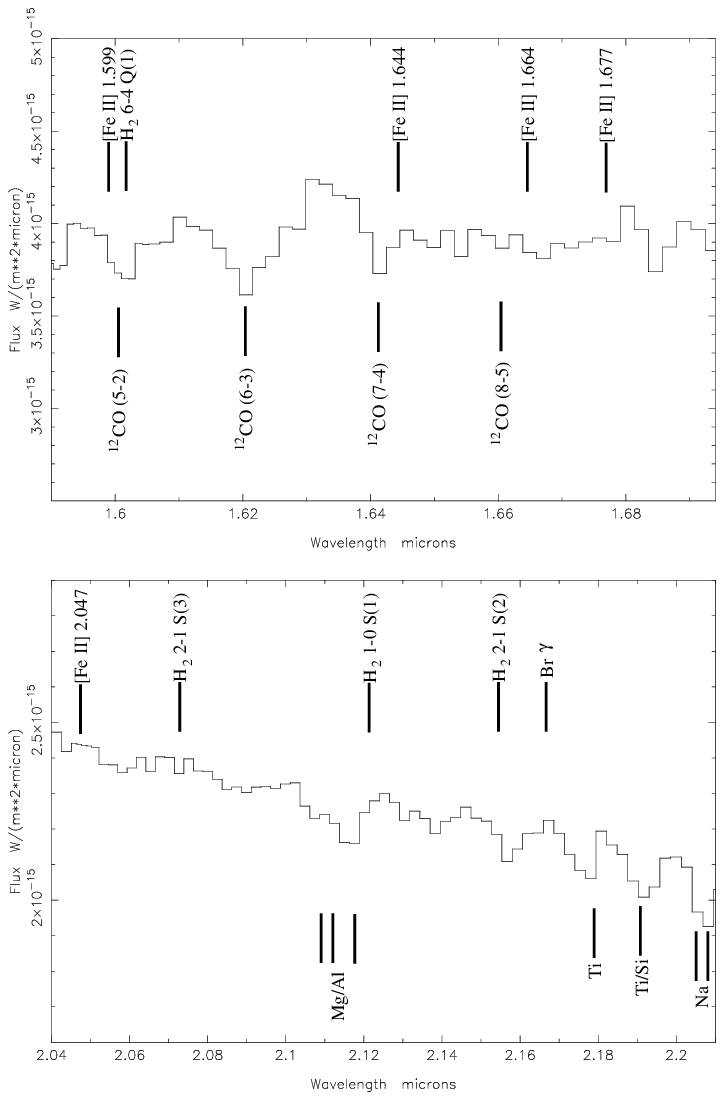,height=13cm}}
\caption{Two regions of the near-infrared spectrum are
shown, shifted to the rest frame.
The expected positions of [Fe {\sc ii}], $^{12}$CO, and H$_{2}$ lines are 
marked. While
some absorption lines are evident, none of the expected emission 
lines are clearly detected. Typically, the strongest lines would be
the [Fe {\sc ii}] lines at $1.644\mu$m and $1.664\mu$m and
the H$_{2} 1-0$\,S(1) line.
The wavelength calibration is accurate to 0.003\,$\mu$m. }
\label{HKspec}
\end{figure*}

In Figure~\ref{HKspec}, the near-infrared HK spectrum has some absorption 
lines, but no emission lines are evident.
Significant lines that lie within the spectral range at this redshift include
two hydrogen recombination lines (Pa$\alpha$ and Br$\gamma$), several
[Fe\,{\sc ii}], and H$_{2}$ lines.
Hydrogen recombination lines are expected to be stronger from starbursts
than AGNs due to the H\,{\sc ii} regions surrounding young hot
OB stars (Hill et al. 1999). This may suggest that there is not a strong contribution
from starburst regions within a radius of 1.6\,arcsec (the offset of the
extracted spectrum) from the nucleus.
The lack of [Fe\,{\sc ii}] and H$_{2}$ lines is less indicative as their
origins can be starburst or AGN. For starburst regions, [Fe\,{\sc ii}] lines 
can be excited in the cooling tails of supernova remnants, or directly
photoionized by young OB stars.
Alternatively, X-rays from the nuclear source may photoionize
the NLR clouds, or radio jets may shock-excite iron. In section~\ref{AGN},
we show that the optical line ratios in the nuclear regions provide
strong evidence for AGN activity while further out, starburst processes are
found. The lack
of [Fe\,{\sc ii}] lines in the nuclear spectrum indicates that the AGN processes are not exciting
iron in this galaxy. There may be a lack of iron within 1.6\,arcsec of the nucleus,
or alternatively, this could point to the starburst processes as more
likely to be the excitation mechanism for iron in active galaxies.

\subsubsection{Rotation Curves}
\label{Rotcurve}

\begin{figure*}
\centerline{\psfig{figure=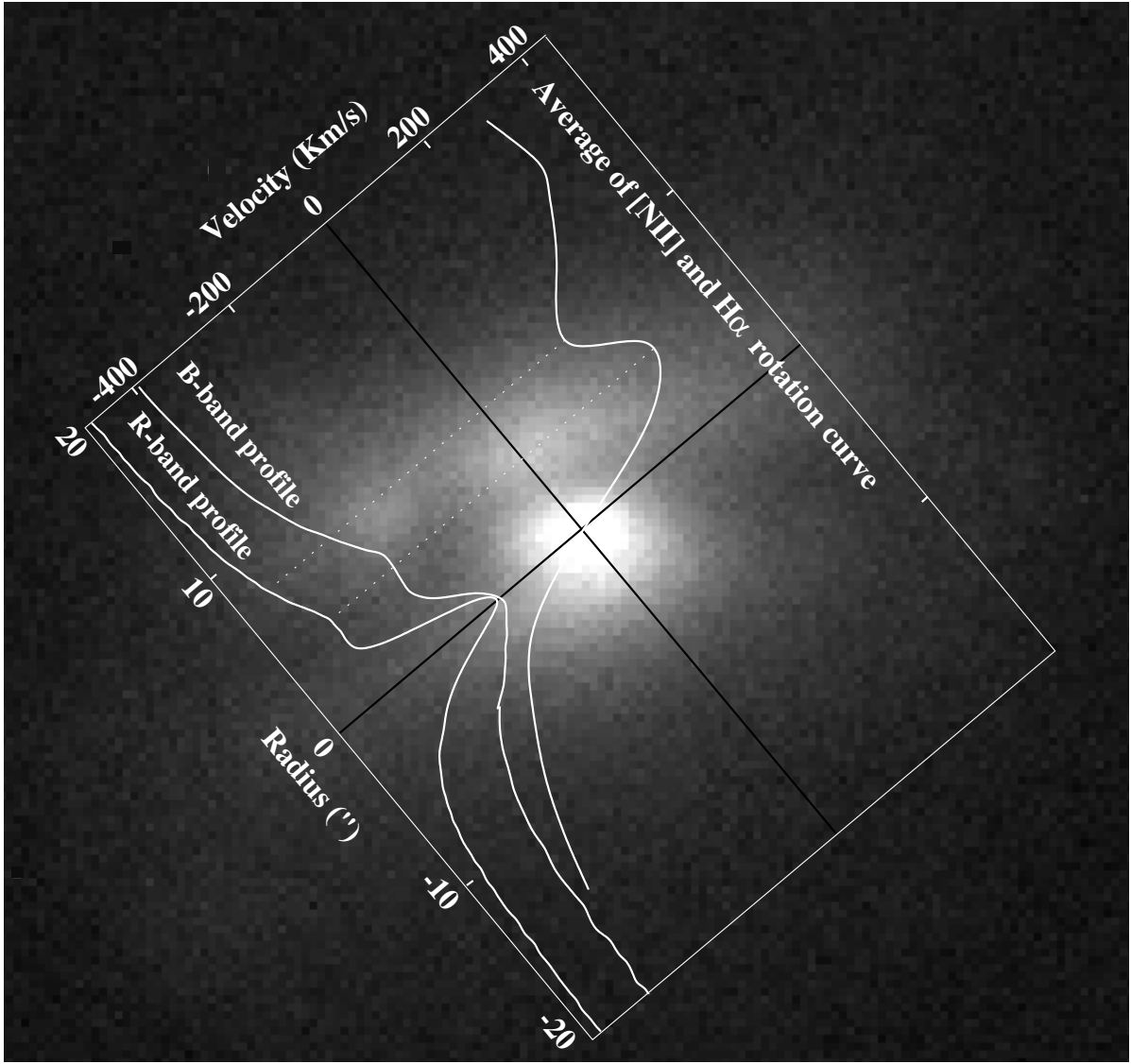,height=15cm}}
\caption{Graph of the H$\alpha$ and [N\,{\sc ii}]$\lambda6584$
average rotation 
curve and the profiles across the $R-$ and $B-$band images along the slit. This
is then overlaid
on the $B-$image such that the graph radius-axis aligns with the slit
position angle through the nucleus.}
\label{cooloverlay}
\end{figure*}

The ionised gas rotation curve in Figure~\ref{rotcurve}(c) clearly shows fast rotation 
along the slit direction, which has a position angle $11^{\circ}$ from the radio axis. The actual rotation
axis is unknown. The south-west side is approaching and the north-east (dust 
lane) side is receding. As the distribution of emission lines is not
symmetric about the nucleus (see section~\ref{spectraobs}), 
the ionised gas extends further to the
north-east than the south west. The rotation curve shows a sudden decrease in velocity
from $5\pm1$\,arcsec to $\sim15$\,arcsec to the north-east. 
A comparison of the features in Figure~\ref{cooloverlay} shows that the dust lane extends to a maximum
offset of $\sim4$\, arcsec from the nucleus, which is before
the decrease in velocity and is therefore unlikely to be related. 
Similarly, Nicholson, Bland-Hawthorn \& Taylor (1992) found 
that the dust lane obscuration has only a minor effect on
an H$\alpha$ image of the dust lane elliptical galaxy, Centaurus A. The 
slowest velocity point in the `dip' does not have a corresponding feature in 
the ESO248-G10 images.

It is noteworthy that the rotation curve shows no sign of a turnover out to 
20 arcsec or $\sim3.3$ effective radii. Extended rotation curves have 
been measured for relatively few ellipticals (e.g. NGC 3115, Capaccioli et al. 
1993) and in each case, have been found to remain flat out to $3-4$ effective radii.

\section{DISCUSSION}
\label{Discussion}

\subsection{Galaxy Rotation and Orientation}

\subsubsection{Orientation of the Principal Axes}
\label{Orientation}
The emission line gas shows fast rotation
along an apparent axis close to the radio axis (Figure~\ref{rotcurve}(c)). 
Elliptical galaxies with rings or disks rotating about an axis
perpendicular to the symmetry plane are not rare. 
A survey of 40 early-type galaxies with
dust lanes by Hawarden et al. (1981), found that polar disk$/$ring galaxies comprised
the largest subgroup, suggesting that they are likely to be stable systems.
Such a polar gas disk is only stable in a triaxial system (Steiman-Cameron \&
Durisen 1982; van Albada et al. 1982). 
Triaxial galaxies have principal axes a, b, and c of lengths A, B, and C
such that A$>$B$>$C, where A=B for oblate
and B=C for prolate ellipticals.
Many attempts have been made to find the orientation of the
three principal axes with respect to the
apparent projection of triaxial galaxies
(van Albada et al. 1982; Steiman-Cameron, Kormendy \& Durisen 1992 and references therein).
The first assumption is that the radio axis is aligned with a principal 
axis of the 
system. It is then commonly accepted that the dust lane will lie perpendicular to
the radio axis for the following reasons: firstly, there is a strong association between galaxies
with radio jets and those with dust lanes, and hence both features are likely
to be produced by a related mechanism such as the gas layer fuelling the
central accretion disk. Secondly, surveys of galaxies with dust lanes and radio jets 
have shown a strong correlation between the jet axis and rotation axis of the
dust lane disk (Kotanyi \& Ekers 1979, Mollenhoff et al. 1992). 

Dust lanes and cylindrical gas disks are thought to be the result of mergers, 
as they are
common features of galaxies with boxy bulges (eg: NGC7332, Fisher,
Illingworth \& Franx 1994), and are often dynamically decoupled from the 
stellar velocity field (Bertola, Buson \& Zeilinger 1992).
Steiman-Cameron \& Durisen (1982) and Steiman-Cameron et al. (1992) 
modelled gas in-falling into a triaxial potential from several merger scenarios.
They found that the accreted gas from the merger of a gas rich galaxy 
and an elliptical would settle into one of the principal planes 
of the resulting
galaxy. The initial orientation of the angular momentum vector of the
accreted gas, relative to the principal planes, will determine into which
plane it settled through differential precession and dissipation.
The angular momentum of the accreted gas 
can be parallel, anti-parallel or perpendicular to the stellar rotation. 
If the figure rotation axis lies in the plane of the dust lane, the 
result is tumbling about this plane.
Furthermore, if the gas and dust are from separate merger or interaction
events, then the cylindrical gas disk rotation can be
perpendicular, parallel or anti-parallel to the dust lane and radio axis. 

Attempts to model stable orbits in triaxial potentials have shown that the only 
stable axes of rotation are the long and the short axes, a and c respectively. 
Rotation about the
intermediate axis is unstable (van Albada et al. 1982).
If the axis of rotation of the dust disk (and hence
the radio axis) is the short axis, then a stable polar gas disk is therefore 
confined to be orbiting the long axis, and vice versa 
(Steiman-Cameron \& Durisen 1982, Merritt \& de Zeeuw 1983). 
The figure rotation
may be parallel or perpendicular to the radio axis but cannot occupy the
intermediate axis. If the figure rotation is perpendicular to the radio axis, 
the triaxial figure is tumbling.
Furthermore, if the surface figure is tumbling,
it must do so much slower than the
orbital period of gas in the polar disk, so the disk will not become unstable.
The apparent stability of polar gas disks
implies that these galaxies are triaxial and that the surface figure 
rotates much slower than the polar gas disk. 

In order to find stable configurations for rotating triaxial systems, it will 
be assumed that the figure rotation is in the same sense as the stellar 
rotation. This applies if the
mass distribution follows the light distribution. If so,
then the four possible stable configurations are given
in Table~\ref{systems} (Merritt \& de Zeeuw 1983 , van Albada et al. 1982).
The warped dust lane in ESO248-G10 is aligned closest to the major axis (Saripalli et al. 1994).
This classifies ESO248-G10 as configuration 3 in Table~\ref{systems}. Hence the radio
axis is the short axis and the figure rotation is about the long axis. 
While the actual gas rotation position angle has not been measured, the 
spectrum showed fast rotation along the slit at a p.a. of $40^{\circ}$. 
The radio axis is only $11^{\circ}$ away at a p.a. of $51^{\circ}$.
Therefore the axis about which the gas orbits is likely to be in, or close to
the plane of the dust lane, perpendicular to the radio axis.
If the gas is from a recent merger it may not yet have settled into a
principal plane.
If the gas orbit has settled and is stable, then the rotation axis is
in the plane of the dust disk and confined to the long axis.
In that case, the stars and gas are tumbling about the long axis with the
figure rotation much slower than that of the gas.

\begin{table*}
\centering
\begin{minipage}{170mm}
\caption{Four possible configurations for rotating triaxial systems.}
\begin{tabular}{@{}lcccc}
\tableline
\tableline
 & Figure Rotation &Rotation of Dust Disk &Form of Dust Lane and & \\
Configuration &(Stellar Rotation)  &(Radio Axis) &  App. Major$/$Minor Axis  &Example \\
\tableline
1  & short axis & short axis &flat - major  & M31 \\
2  & short axis & long axis & warped - minor & CenA, M84?\\
3  & long axis & short axis & warped - major & M84?, ESO248-G10\\
4  & long axis & long axis & flat - minor & \\
\tableline
\end{tabular}
\label{systems}
\end{minipage}
\end{table*}

\begin{figure*}
\centerline{\psfig{figure=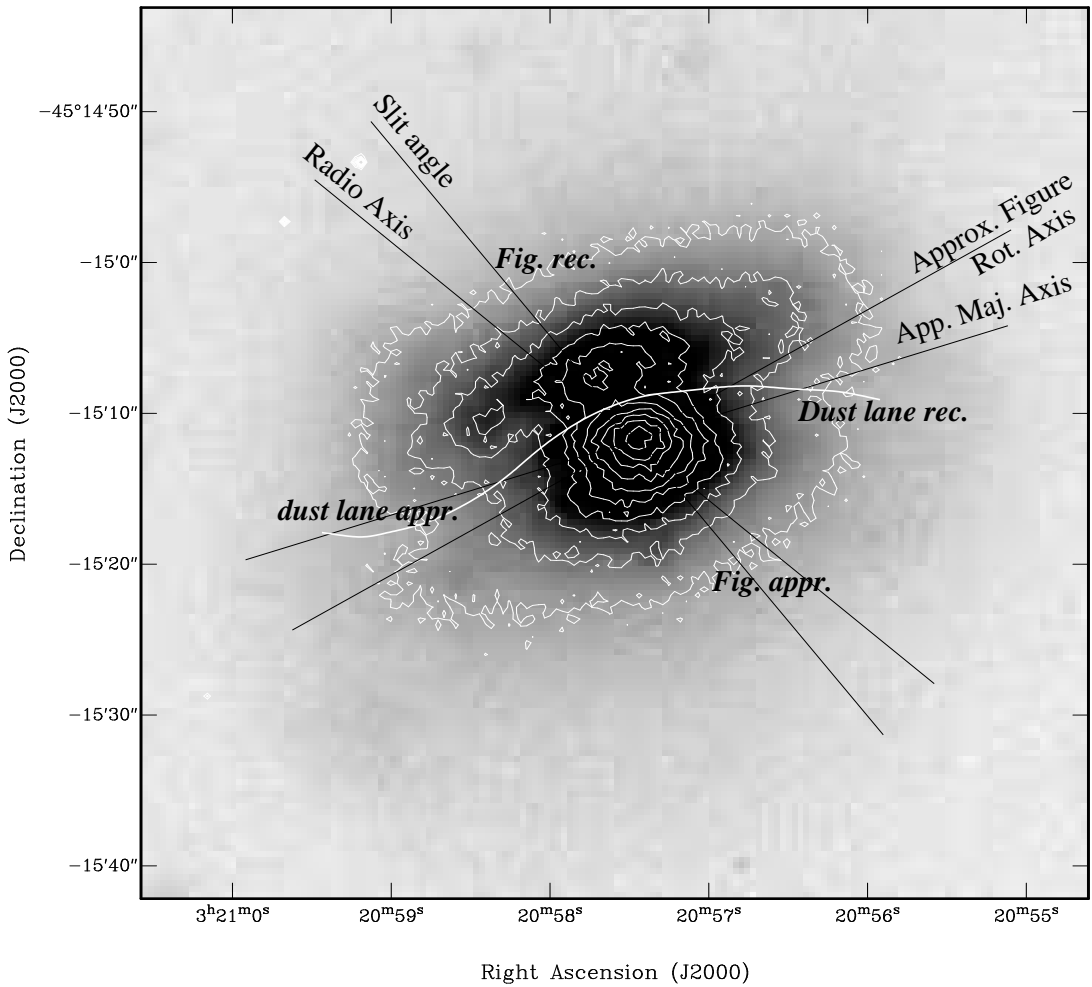,height=15cm}}
\caption{Orientation of observed axes overlaid on the $B-$band image and contours. }
\label{angles}
\end{figure*}

It has been argued that warped dust disks are transient and have not yet 
settled into their rotation plane. However, 
van Albada et al. (1982) showed that slow figure rotation or tumbling 
about an axis in the plane of the dust disk can result in a stable warped 
disk due to the Coriolis force. The resulting warp increases with radius.
Furthermore, if the warped dust lane is along the major axis, then
the rotation of the warp is prograde to the 
figure rotation when viewed down the figure rotation axis (Merritt \& De Zeeuw
1983, Kormendy \& Djorgovski 1989). 
While the gas and stars are both rotating about the long axis, the sense
of the stellar rotation is unknown. It would be plausible for both to
rotate in the same sense, although it is not impossible to find 
counter-rotation about the same axis (eg. NGC7332, Fisher et al. 1994).
If the gas and stars in ESO248-G10 are rotating in the same sense, then its 
figure rotation is away from the observer to the north-east. 
A stable system would therefore imply that the south-east dust lane warp is 
rotating away from 
the observer.  Furthermore, the optical images reveal the strongest extinction 
from the dust lane is north of the nucleus (Figure~\ref{images}(a)) and hence the dust lane tilts away 
from the sky plane to the north-east. As the warp in the dust disk is expected to
increase with radius, this confines the long (figure rotation) axis to have a
position angle
greater than $107^{\circ}$ (from the inner dust lane). Furthermore,
the projected long axis is likely to be close to perpendicular 
(p.a.\,$=130^{\circ}$) to the slit orientation as the measured velocity is so 
fast. Figure~\ref{angles} shows the orientation of the observed axes.
The long figure rotation axis is shown at a nominal p.a. of $119^{\circ}$,
midway between the above values.

\subsubsection{A Model for the Orientation of ESO248-G10}
\label{model}
  
\begin{figure*}
\centerline{\psfig{figure=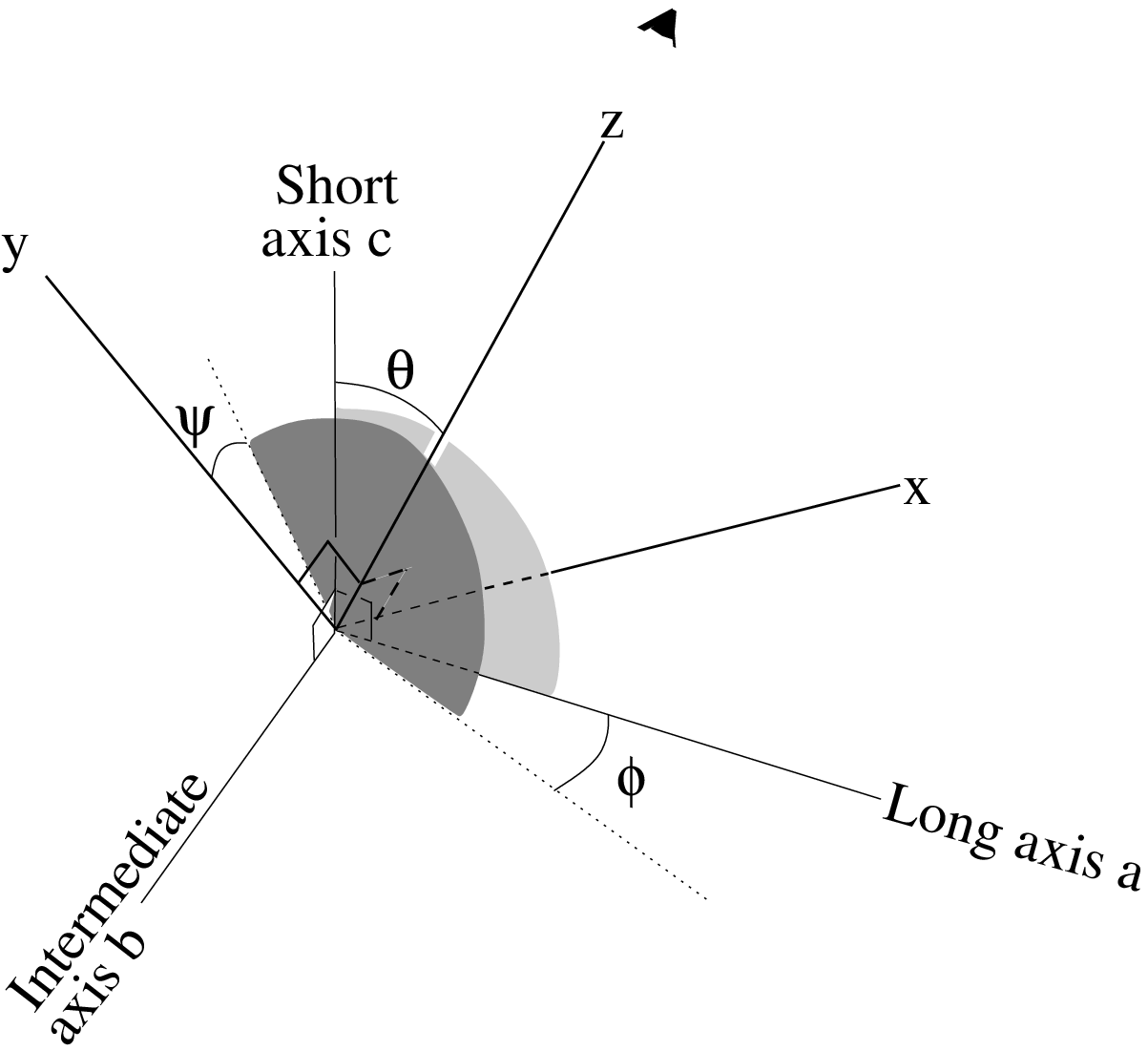,height=9cm}}
\caption{Viewing angles $\theta$ and $\phi$ represent the angles between
the principal axes, a, b, and c, and the viewing axes, x, y, and z, as shown. The z
axis is chosen to be along the line of sight, putting the x and y axes in
the sky plane. $\psi$ is apparent angle between the projection of the minor 
axis (y) and the short principal axis, c, on the sky. }
\label{0319orie}
\end{figure*}

The radio axis is expected to align with the long or the short principal axis (see
section~\ref{Orientation}). However, due to orientation effects, the radio 
axis does not appear coincident with the apparent major or minor axes.
In such a triaxial system, to determine the orientation of the principal axes 
and their relative lengths requires five parameters. Firstly, we define 
{\it $\psi$} as the apparent angle between the minor axis and the projection
of the short principal axis on the sky. 
Two more parameters define the principal axes lengths such that $\zeta=B/A$
and $\xi=C/A$.
The remaining viewing angles are given by the Euler angles 
({\it $\theta$},{\it $\phi$}). We define two coordinate systems in 
Figure~\ref{0319orie}: (a,b,c) are
the axes associated with a triaxial ellipsoid, while (x,y,z) define the
sky plane with z along our line of sight. $\theta$ is the angle of axis c
to the line of sight. The angle from axis a to the intersection of the c$-$z 
and a$-$b planes defines $\phi$. 

The relationships among these five parameters have been extensively discussed 
in the
literature since Stark (1977) derived the apparent brightness distribution of a
set of similar triaxial ellipsoids (see also van Albada et al. 1982; 
de Zeeuw \& Franx 1989). We extend the notation and equations of 
Hui et al. (1995) to the case of ESO248-G10, where the radio axis is along the 
short axis, and $\psi$ is $34^{\circ}$ (Saripalli et al. 1994).
It can be shown that
\begin{equation} 
\xi=\sqrt{0.404\,L\,{\rm cot\theta\,cosec\theta}\,-\,M\,{\rm cot^{2}\theta}\,+\,N\,{\rm cosec^{2}\theta}},
\label{eq.h}
\end{equation}
where
\begin{eqnarray}
M & = & (\zeta^{2}\,{\rm sin^{2}\phi}\,+\,{\rm cos^{2}\phi}), \\ 
N & = & ({\rm sin^{2}\phi}\,+\,\zeta^{2}\,{\rm cos^{2}\phi}), \\ 
L & = & {\rm sin2\phi}\,(\zeta^{2}-1)  
\end{eqnarray}
and $\psi$ is the apparent angle of the short axis to the minor axis as defined above, when
\begin{equation}
M\,{\rm cos^{2}\theta}\,+\xi^{2}\,{\rm sin^{2}\theta}\,-\,N\,+\,2.475\,L\,{\rm cos\theta}\,\leq\,0.
\label{eq.limit}
\end{equation}

\begin{figure*}
\centerline{\psfig{figure=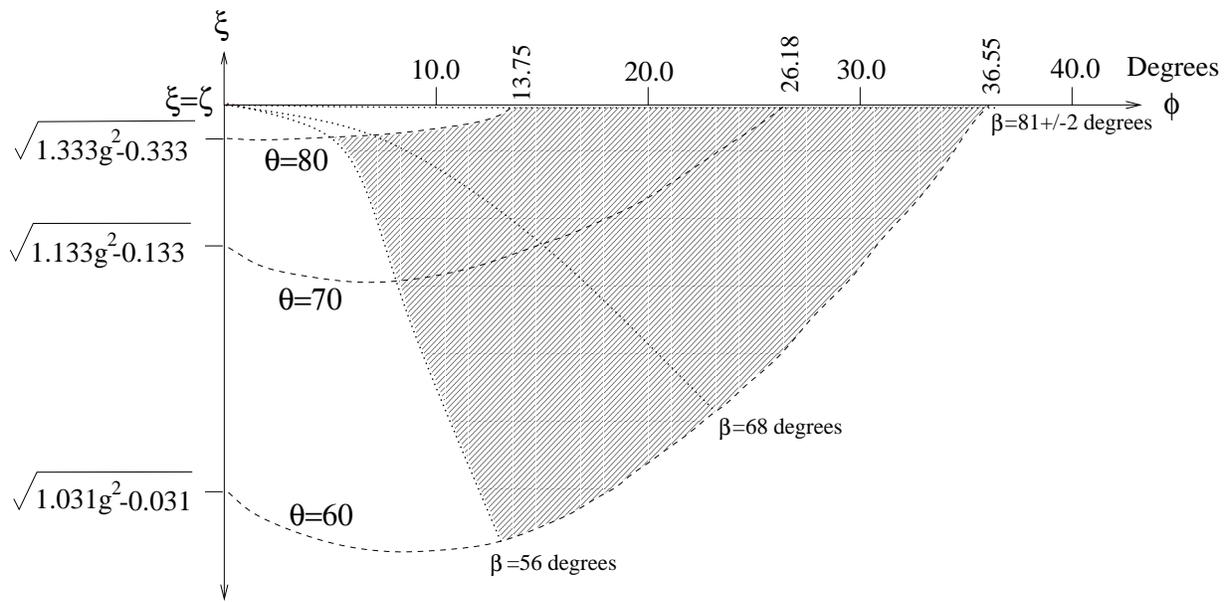,height=8cm}}
\caption{$\xi$ vs $\phi$. The shaded region represents the range of values
allowed by equation~\ref{eq.h} and the constraint $\zeta>\xi$, for the chosen
range of $\theta$. }
\label{constraint}
\end{figure*}

We now set about finding values of $\theta$, $\phi$, $\xi$, and $\zeta$
that satisfy equations~\ref{eq.h}$-$~\ref{eq.limit}.
To constrain these four parameters, several assumptions will be made. 
As the radio structure appears so large (Saripalli et al. 1994), we assume 
that the radio axis (and hence the short axis) is close to the sky plane,
say, $\theta\,>\,60^{\circ}$. However, with the radio axis perpendicular
to the dust lane (see section~\ref{intro}), the tilt of the dust lane suggests the
radio axis is projected away from the sky plane to the north-east, and hence
$\theta\,<\,90^{\circ}$. We therefore adopt $60^{\circ}\,<\,\theta\,<\,80^{\circ}$.
There are then two constraints on $\phi$. Firstly, for a given $\theta$,
$\phi$ must be chosen to give the correct apparent angle between the figure rotation
(long) and radio (short) axes, called $\beta$. From section~\ref{Orientation},
the figure rotation axis has p.a.\,$>107^{\circ}$ and the radio axis has p.a.\,$=51^{\circ}$, therefore $\beta>56^{\circ}$. 
Secondly, as $b>c$, equation~\ref{eq.h} requires that
for any given $\theta$, $\phi$ is limited by $\zeta>\xi$. 
Figure~\ref{constraint} shows the set of curves for which $\zeta>\xi$. For
any chosen $\theta$, there is a maximum possible $\phi$. 
This restricts the range of values
to those shaded in Figure~\ref{constraint}.
Modelling has shown
that choosing the maximum $\phi$ value puts the figure rotation axis 
approximately perpendicular to the slit at p.a. of $132\pm2^{\circ}$, and 
$\beta=81\pm2^{\circ}$.
We can therefore constrain the figure rotation (long) axis to be between 
$107^{\circ}$ and $132^{\circ}$; we (arbitrarily)
adopt the midpoint of
$119^{\circ}$, which gives $\beta=68^{\circ}$. 

The dust lane lies in the plane perpendicular to the radio axis (see section~\ref{Orientation}). 
$\theta$ can be chosen to reproduce the apparent tilt of the dust lane in 
the perpendicular plane. The warp in the dust lane makes the tilt difficult
to judge, but within the range $60^{\circ}<\theta<80^{\circ}$, modelling suggests $\theta=65^{\circ}$.
For $\beta=68^{\circ}$, $\phi$ is found to be $19\pm1^{\circ}$.

\begin{figure*}
\centerline{\psfig{figure=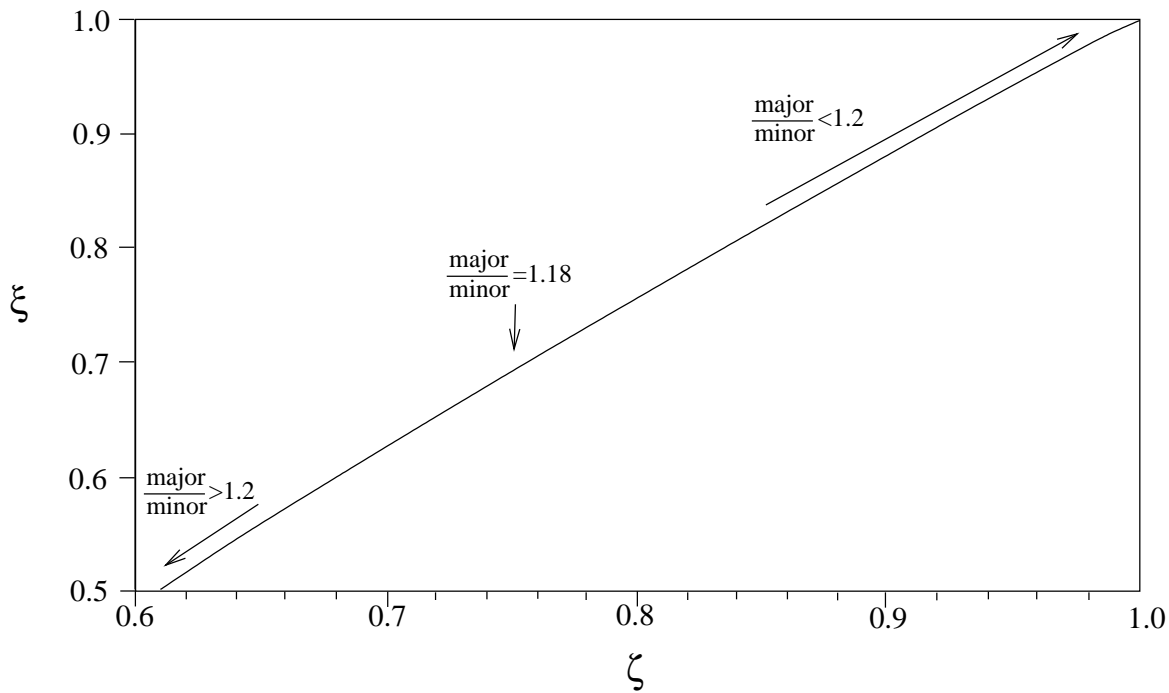,height=9cm}}
\caption{$\xi$ vs $\zeta$ for the chosen values of $\theta$, $\psi$, and $\phi$.
The apparent major-to-minor axis ratio is dependent on $\xi$ and $\zeta$
as shown.  }
\label{hvsg}
\end{figure*}

Once the angles are decided, the dependence of the axis ratios is shown
in Figure~\ref{hvsg}.
The intrinsic ratio $C/A\geq1/2$ is implied by the small
number of radio galaxies flatter than E5
(Hummel 1980, referenced in van Albada et al. 1982). Therefore $0.5<\xi<1$ and $\zeta>\xi$.
The choice of $\zeta$ and $\xi$ affect the apparent photometric major$/$minor
axes ratio. Saripalli et al (1994) found this axis ratio to be 1.18.
Modelling an ellipsoid with $\zeta=0.75\pm.05$ and $\xi=0.69\pm.06$ at the
chosen angles was found to reproduce this major$/$minor axis ratio, giving
$1.18\pm0.01$.

\begin{figure*}
\centerline{\psfig{figure=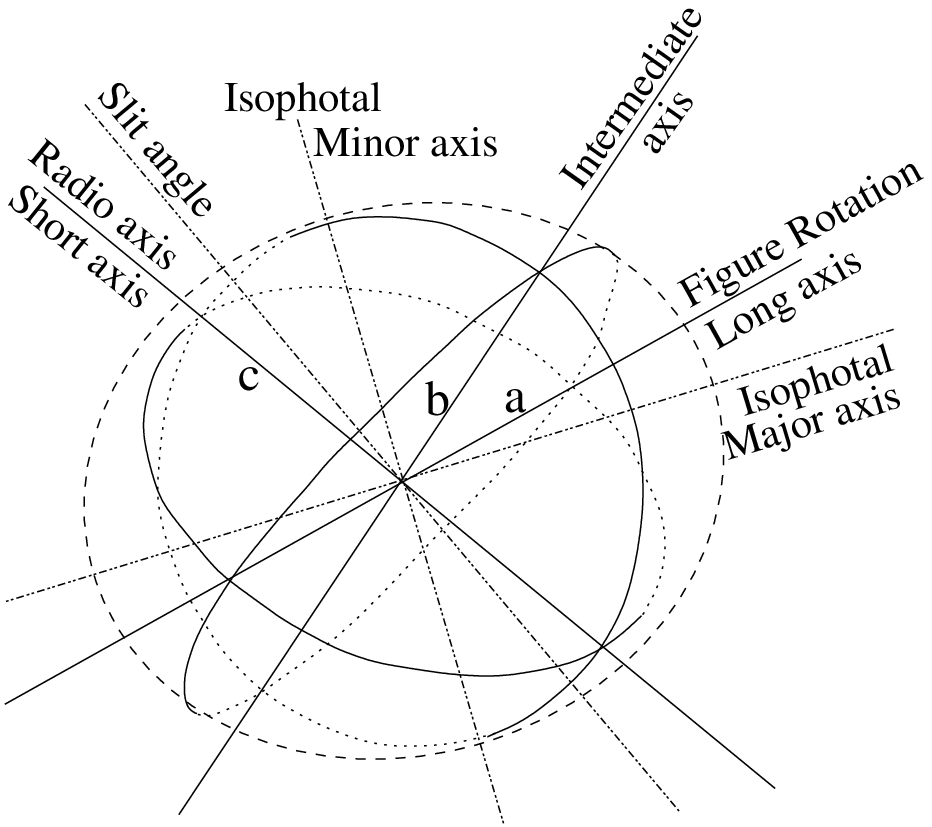,height=9cm}}
\caption{A model for the orientation of ESO248-G10.
The angles are $\theta=65^{\circ}$, $\psi=34^{\circ}$, $\phi=19^{\circ}$, and 
axis ratios of B$/$A\,$=0.75$ and C$/$A\,$=0.69$.}
\label{Fmodel}
\end{figure*}

Figure~\ref{Fmodel} shows our adopted model for ESO248-G10 with $\psi=34^{\circ}$, $\theta=65^{\circ}$,
$\phi=19^{\circ}$, $\zeta=0.75$, and $\xi=0.69$. While not the only
possible model, it does represent a self consistent choice of angles
and it reproduces the observed features.

\subsection{Nature of the Nuclear Source and Circumnuclear Structure}
 
\subsubsection{Reddening Processes}     
\label{reddening}
\begin{figure*}
\centerline{\psfig{figure=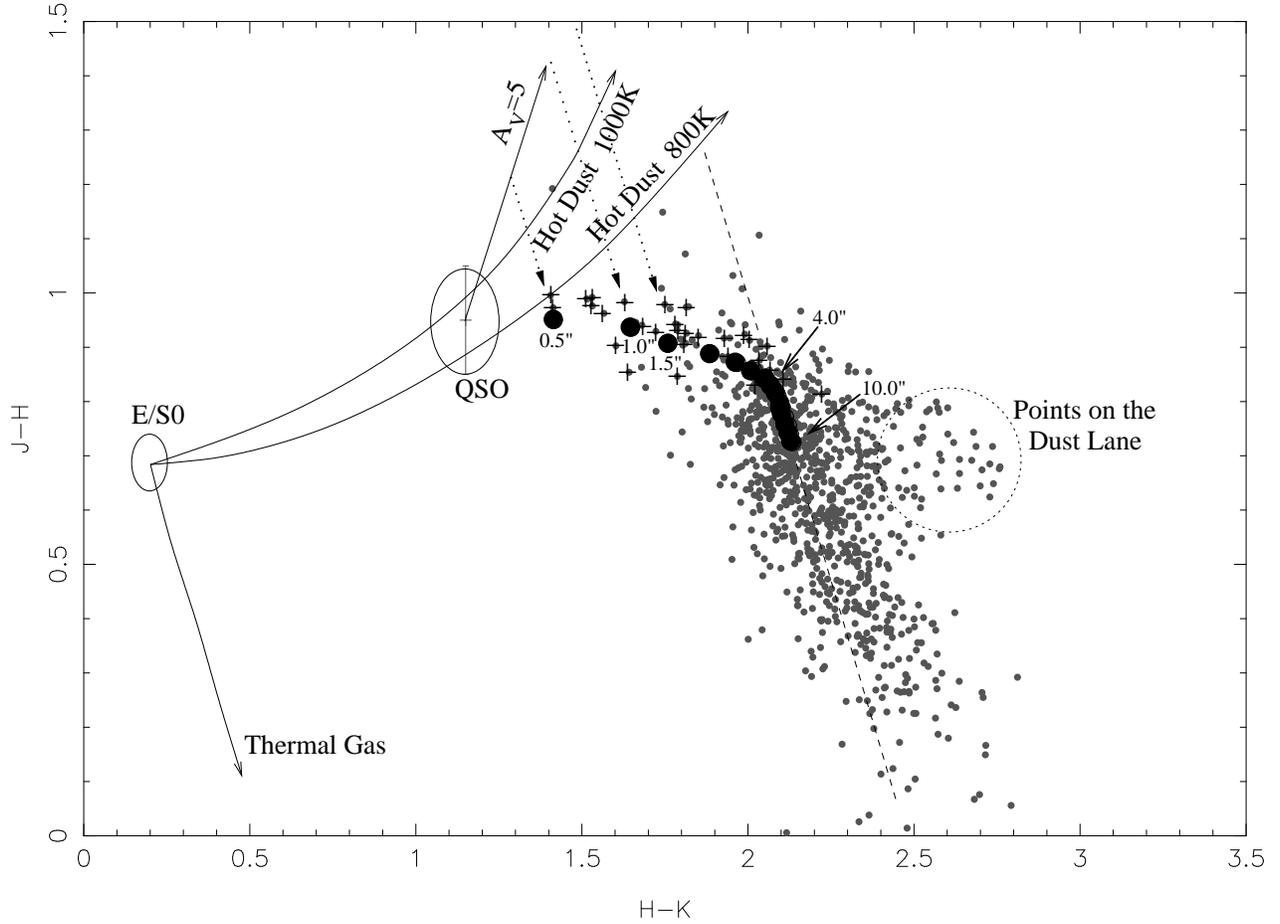,height=13cm}}
\caption{$J-H$ vs $H-Kn$ for ESO248-G10. Typical elliptical galaxy and QSO colours are
marked by the two ellipses. An extinction vector of
A$_{V}=5$ is shown along with hot dust vectors for 800K and 1000K. The colours
of $31\times31$\,pixels ($15.5\times15.5$\,arcsec) centred on the nucleus are 
shown by the small dots. The crosses are the 30 pixels closest to the nucleus.
Aperture photometry colours marked by large dots represent apertures
of 0.5, 1, 1.5, 2, 2.5, 3, 3.5, 4, 4.5, 5, 6, 7, 8, 9, 10\,arcsec from left to right.
For clarity only some of these aperture radii have been labelled.
Points within the dotted circle all lie on the dust lane. The dashed line is
parallel to the thermal gas vector. Dotted lines indicate a themal gas vector which could
combine with the dust extinction or hot dust vectors to give the observed aperture photometry
points. }
\label{2colplot}
\end{figure*}

The extinction images in Figure~\ref{images}(d) reveal significant
reddening around the nucleus. Thermal gas emission, hot dust emission and
dust extinction can all affect the galaxy colours. To distinguish between these
processes, the two-colour pixel plot in Figure~\ref{2colplot} was
formed from the values of $31\times31$\,pixels ($15.5\times15.5$\,arcsec) 
around the nucleus; the typical colours of E$/$S0 galaxies and
QSOs are shown on the figure. Also shown are vectors representing 5 magnitudes 
of visual extinction, 
hot dust at temperatures of 800 and 1000K mixed with a late-type stellar 
population, and thermal gas comprised of hydrogen and helium continuum 
and emission lines mixed with late-type stars (Alonso-Herrero et al. 1998).
Aperture photometry colours for a range of increasing apertures are also plotted
in Figure~\ref{2colplot}.

Notice that in Figure~\ref{2colplot} both the individual points coincident with the nucleus and the 
smallest aperture photometry point are bluer than points at larger radii.
With increasing radius, the colours
initially become redder along an almost horizontal vector. Beyond a
radius of 4\,arcsec, the points then follow
the direction of the hot gas vector. Two scenarios could explain this
pattern. The colours may be the sum of a hot dust and a hot 
gas vector.  As the radius increases, the contribution 
of the hot dust and hot gas both grow
until, at 5 arcsec, the hot gas processes become
dominant and there is no further contribution from hot dust.
Similarly, dust extinction, not emission, could contribute to
reddening out to 5 arcsec. From the two-colour pixel plot alone, it is not
possible to distinguish between these two scenarios and
we now examine in turn the evidence for dust extinction and hot dust 
emission.

\noindent
(a){\it Dust extinction}: This is plausible since
the optical images clearly show a dust lane. The thickest portion of 
the dust lane does not appear to obscure our view of the nucleus,
as the tilt results in the foreground dust passing above the nucleus. Two
observations support the interpretation of the
two-colour pixel plot as indicating dust extinction plus thermal gas.
Firstly, the nuclear pixels do
not show as much dust extinction as those at larger radius. Secondly,
the aperture photometry colours follow a dust extinction plus thermal 
gas emission vector until the 4\,arcsec aperture point.  This corresponds to the
distance to which the dust lane extends from the nucleus in the direction
where the dust lane is closest to the nucleus 
in the optical images.
Beyond 4\,arcsec, the aperture photometry points follow the thermal gas
vector and show no further contribution from dust extinction.
Furthermore, the individual points within the dotted circle in Figure~\ref{2colplot}, 
which have the highest reddening, are 
the pixels on the dust lane; dust extinction provides a logical explanation
for these points. If so, the extinction to the dust lane pixels
would be around 13 visual magnitudes.

\noindent
(b){\it Hot dust emission}:
The second explanation of the two-colour pixel plot involves hot dust emission
as the prominent reddening source out to 4\,arcsec.
Four mechanisms may contribute to hot dust emission: 
(i)transient heating of small grains by reprocessed X-rays, 
(ii)scattered nuclear radiation, 
(iii)heating by UV photons from the central source, 
or (iv)in situ heating of dust by hot stars. 
We now examine each mechanism in turn.
The first option cannot be ruled out in this case, as models
of X-ray heating of small grains are inconclusive. The processes of grain
destruction and heating are not well enough understood to produce consistent
models of this process. For the second mechanism, hot dust emission close to the central
source can be scattered into the line of sight by clouds further from
the nucleus. A polarisation image would be required to prove scattering.
However, models of similar galaxies
have shown that essentially all the emission would need to be scattered
towards the line of sight (Alonso-Herrero et al. 1998). Hence, 
this scattering scenario is unlikely.
The third mechanism for hot dust emission
is heating by UV photons from the 
central source which will reradiate in the near infrared. 
UV heating has been considered to be a major source 
of near-infrared emission from regions close to the nuclei of Seyfert galaxies 
(Alonso-Herrero, Ward \& Kotilainen 1996; Kotilainen et al 1992; Bryant \& Hunstead 
1999). 
If the same process is responsible for extended thermal dust, then it
must be shown that UV photons from the central source can heat the dust
to the observed temperatures out to the radius required. 
For a dust optical depth of $\tau_{UV}$, a total
UV luminosity of L$_{UV}$ (in units of $10^{46}$\,erg\,s$^{-1}$) would heat the dust
to a temperature T (K) out to a radius r (pc) given by,
\begin{equation}
T=1650(\frac{L_{UV}}{r^{2}}.e^{-\tau_{UV}})^{\frac{1}{5.6}} 
\end{equation}
(Barvainis 1987).
The near-infrared colours indicate that if hot dust is the reddening mechanism, then this
dust extends to $\sim6.5$\,kpc ($\sim4$\,arcsec) before hot gas becomes the 
dominant 
mechanism. The typical temperature for such hot dust is around 800K. A higher
temperature would produce a steeper vector on the 2-colour plot.
At 1500K the dust would be evaporated by the UV radiation (Barvainis 1987).
Therefore, to heat dust to a temperature of 800K out to a radius of 6.5\,kpc
would require a UV luminosity of $1.5\times10^{51}$\,erg\,s$^{-1}$, three orders
of magnitude greater than the 
most luminous quasars.
Therefore, it is very unlikely that UV radiation from the central source
could be responsible for heating dust to such a radius.
The fourth origin of hot dust emission requires groups of hot stars to
provide local heating of the dust far from the central source.
Massive star formation in the galaxy halo produces UV emission which can
be almost entirely converted to thermal dust radiation (Alonso-Herrero et al.
1998).
Not only can star forming regions heat the dust, they can
also heat the gas. As thermal gas emission is an important
processes further from the nucleus, the evidence for
starburst regions will be discussed at length in the next section.
The star forming regions will be shown to extend to
the radius of the proposed hot dust emission.

\subsubsection{Distribution of AGN and Starburst Processes}
\label{AGN}

Table~\ref{fluxratios} compares three diagnostic emission line ratios 
measured in 3 pixels ($2.3$\,arcsec) centred on the nucleus with
the expected ratios for
starbursts and AGNs (Radovich \& Rafanelli 1996, Hill et al. 1999). 
For each diagnostic, the nucleus is clearly identified with AGN emission.

\begin{table*}
\centering
\begin{minipage}{120mm}
\caption{Measured emission line ratios for three diagnostic emission
line pairs, measured at the nucleus. }
\begin{tabular}{@{}lccc}
\tableline
\tableline
Line Ratio & Measured value &Value for pure starburst &Value for AGN  \\
     &  & &   \\
\tableline
{\large $\frac{\rm{[N\,{\sc ii}]}\lambda\,6584}{\rm{H}\alpha}$}   & $2.2\pm0.2$  & $\leq\,0.6$  & $\geq\,1.2$\\
 & & & \\
{\large $\frac{\rm{[S\,{\sc ii}]}\lambda\,6716+6731}{\rm{H}\alpha}$}   & $0.73\pm0.2$  & $\leq\,0.25$  & $\geq\,0.32$ \\
 & & & \\
{\large $\frac{\rm{[O\,{\sc i}]}\lambda\,6300}{\rm{H}\alpha}$}   & $0.18\pm0.05$  & $\leq\,0.05$  & $\geq\,0.1$ \\
 & & & \\
\tableline
\end{tabular}
\label{fluxratios}
\end{minipage}
\end{table*}

In Figure~\ref{rotcurve}(d), horizontal lines represent five regions where
pixels were extracted to make spectra. The corresponding spectra and
the [N\,{\sc ii}]$\lambda6584/\rm{H}\alpha$ flux 
ratios are shown below the rotation curve.
For radii less than $\sim5$\,arcsec from the nucleus, the ratio is $>1.2$ as
required for AGNs. However, beyond this radius, the ratios drop to the
range expected for starburst$/$composite regions ($\leq0.8$). Therefore, star 
formation regions are found beyond the radius at which the slow rotation `dip'
(see section~\ref{Rotcurve})
starts on the rotation curve. It is notable that whereas the slow
rotation `dip' is only seen to the north-east of the nucleus, the evidence of star formation
begins at the same radius to both the north-east and south-west.

\subsubsection{North-east `Dip' in Rotation Curves}
\label{NEdip}

Rotation curve features similar to the north-east velocity `dip' have been observed
in other galaxies. In each case, however, the explanation for the dip does not 
appear relevant to ESO248-G10. 
Ionised gas rotation curves with a steep gradient followed by a dip then
a shallower gradient have in some cases (eg: NGC253 Arnaboldi et al. 1995)
been attributed to a nuclear ring. 
Capaccioli \& Longo (1994) suggest that stellar rotation curves with a sharp rise
up to a maximum followed by a dip then a shallower rise, are typical of
lenticular galaxies (eg: NGC1553, an S0 galaxy), in which the dip marks the transition 
between disk-dominated and bulge-dominated parts of the galaxy. 
However, although the ESO248-G10 rotation curve shows a dip to the 
north-east, the lack of a corresponding dip to the south-west
appears to rule out a nuclear ring or lenticular galaxy features.
We now examine alternative models which include a second velocity system to the
north-east.

\subsection{Possible Models of the Host Galaxy}
\label{scenario}

\begin{figure*}
\centerline{\psfig{figure=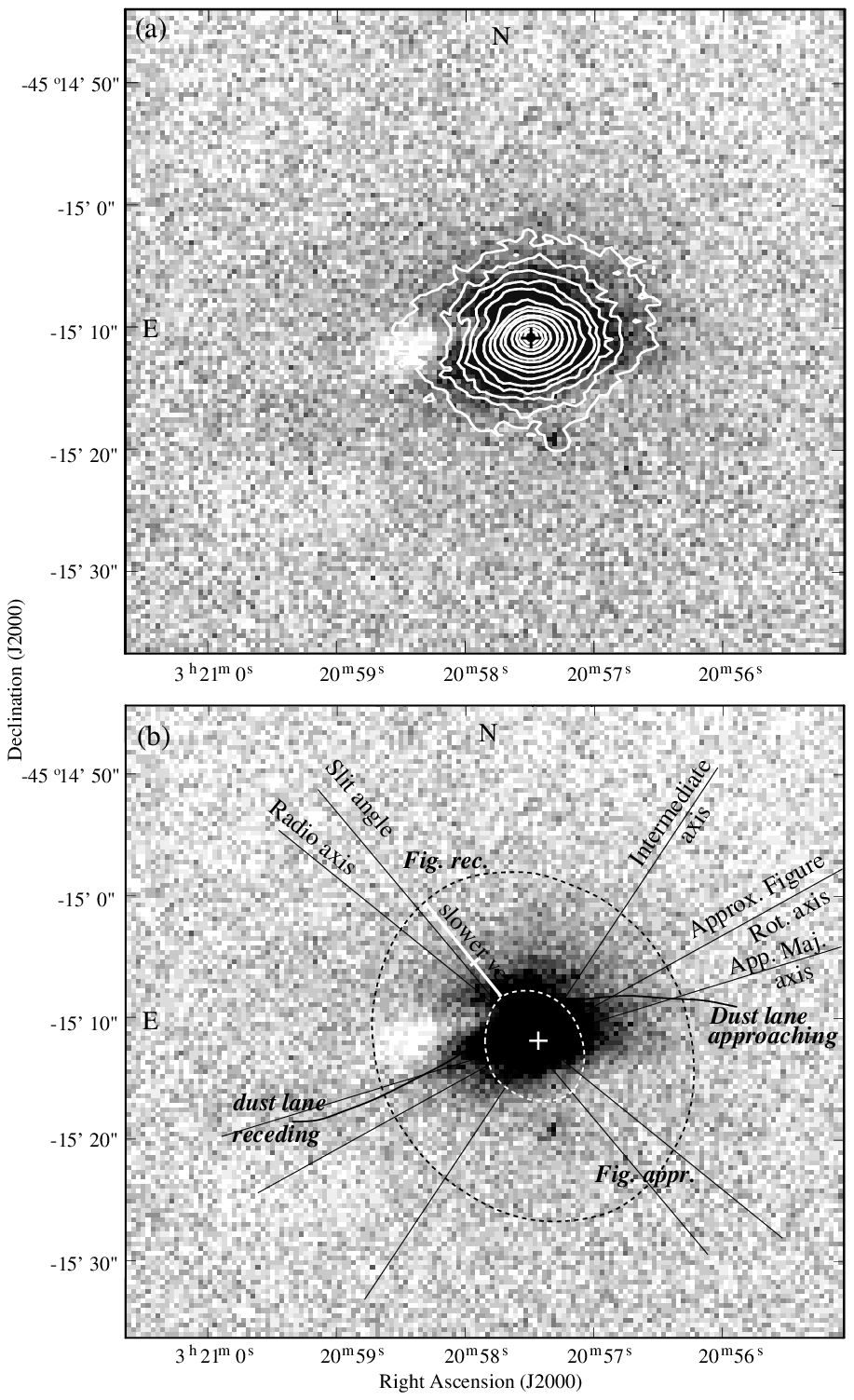,height=18cm}}
\caption{(a)$B-R$ image overlaid with J-band contours. The nucleus is marked by the cross. 
Darker regions represent redder colours.  There is clearly a blue region to the east of the nucleus. 
Contour levels are 17.9, 18.1, 18.4, 18.7, 19.1, 19.3, 19.7, 19.9, 20.1, 20.4, 20.9, 21.2, 21.6 mag\,pixels$^{-1}$. The J-band has distorted contours along the westward
extension 
of the blue patch to the west.  This is unlikely to be due to the dust lane as 
it is not apparent in the contours on the west side of the galaxy. 
(b) $B-R$ image with axes marked. The white
line along the slit position angle, defines the radii (5-15\,arcsec) of the slower 
velocity `dip' from the rotation curve. The local minimum velocity in the
`dip' is marked by the perpendicular dash across the white
line. Two orbits about the figure rotation axis are shown by the 
dashed lines, with radii chosen to lie at either end of the blue feature. 
Rotation in this sense would map the blue feature onto
the region of slower velocity (see section~\protect\ref{scenario}). }
\label{R-B}
\end{figure*}

A common source of extended
emission line gas in ellipticals is capture from a merger or interaction with a
gas rich galaxy (see section~\ref{Orientation}). 
In Figure~\ref{R-B}(a) the $R$ image has been divided by the $B$, to give
an $B-R$ image in which darker regions correspond to redder colours.
The nucleus has $B-R=2.5$ and is marked by a cross. 
Approximately 10\,arcsec to the east of the
nucleus is a bluer patch with $B-R=1.3$,
first identified by Saripalli et al. (1994).
This could be a merging companion or a region of blue stars. 
The blue patch appears to be extended towards the west.
$J-$band contours overlaid on the $B-R$ image appear
unaffected by the dust lane to the north-west of the nucleus. It is therefore likely
that the distortion of the $J-$contours to the east, whilst coincident with the
dust lane, is due to the extended emission associated with the blue patch rather
than dust.

If the blue region in ESO248-G10 can be identified with an interacting companion, then
its orbit will be settling into the principal plane closest to the angular
momentum vector of the merging galaxy (see section~\ref{Orientation}).
In Figure~\ref{R-B}(b), the p.a. of the blue patch from
the galaxy nucleus is between the figure rotation axis and the radio axis.
In fact, it could be closer to either. If the merging galaxy is in the 
long$/$intermediate plane, and settling into an orbit about the radio axis, then 
the current merger could be responsible for the warped dust lane. While that 
hypothesis can not be ruled out, it would not support the
distribution of starburst regions or the rotation curve features observed.

An alternative picture involves the merging galaxy being located in the short$/$intermediate
plane and hence settling into rotation with the figure and gas orbits.
The dashed lines in Figure~\ref{R-B}(b) are two orbits about
the figure rotation axis, projected onto the sky plane. Their radii were chosen to
lie at either end of the putative merging galaxy. 
In Figure~\ref{rotcurve}(c), a region of slower velocity was evident from a
radius of $\sim5-15$\,arcsec to the north-east,
with a local minimum
at a radius of $\sim8$\,arcsec from the nucleus.
A white line along the slit direction in Figure~\ref{R-B}(b) marks the radii of 
the `dip', with the cross line at the local minimum velocity. 
If gas is being accreted from the blue galaxy into an orbit about the figure axis,
the radii of this gas when rotated to the apparent slit angle, 
corresponds well with the slower gas region. However, it is likely that the
merging galaxy would have an orbit close to the one shown, but not yet settled 
into the short$/$intermediate plane.
Nevertheless, if the blue region is a merging companion,
then it could be the second velocity system affecting the rotation curve.
This picture suggests
the inner radius of the proposed merging galaxy's orbit is currently $\sim5$\,arcsec. 
Hence the radius at which thermal gas is accreted into the orbit should
be the same either side of the nucleus. Newly accreted gas has not yet merged
with the orbiting gas and could lead to a second
velocity system near the merging galaxy.
Therefore only the north-east of the rotation curve is affected by a second
velocity system.
The gas is heated by starforming regions triggered by the merger.
The detection of starburst emission line ratios should therefore be symmetric about the nucleus.
This symmetry was indeed observed in Figure~\ref{rotcurve}(d), where starburst
line ratios became
dominant over AGN processes beyond $\sim5$\,arcsec on both sides of the
nucleus. 

An alternative explanation could involve the gas originating from a previous
merger, and the blue region as a cluster of stars. This, however, would leave
no evidence in any of the images, for the source of the second velocity
system.
Interpretation of the blue region as merging galaxy provides a self-consistent
explanation of the observed gas rotation, starburst distribution, and optical colours.

\section{FUTURE WORK}

Optical images in better seeing would help to resolve the
dust lane features and the proposed merging galaxy.
Observations to test the interpretation of the blue region as a merging galaxy 
are currently planned, including spectral mapping 
to measure the dynamics and distribution of gas
across the galaxy. The resulting spectra across the dust lane may confirm 
the sense of the dust rotation.
Furthermore, the rotation information across the galaxy
would refine the axis of gas rotation and our model for the
orientation of ESO248-G10.

The radio axis position angle was assumed to be that of the radio jets
far from the nucleus as mapped by Saripalli et al. (1994). 
However, there is evidence that the jet direction may be changing 
systematically with time, possibly indicating precession, and hence 
may not be at exactly the
same position angle as further out. Therefore, a detailed radio map of the
core regions at arcsec resolution could constrain the local radio jet position
angle and hence further constrain the angles found from modelling the system.

In section~\ref{EmissionL} it was shown that there is no [Fe\,{\sc ii}]
within 1.6\,arcsec of the nucleus, where AGN processes dominate.
This may indicate that if iron is present and not excited by the AGN, then
in general, iron excitation may be more likely from starburst
processes: in cooling tails of SNRs or photoionized by young OB stars. 
An [Fe\,{\sc ii}] image and a higher resolution radio image of ESO248-G10 
could be used to test the excitation mechanism for [Fe\,{\sc ii}] in active 
galaxies. 
The emission should be more prominent at a radius $>$\,5 arcsec 
(see section~\ref{AGN}) if it is due to starburst processes. Furthermore,
a correlation between the line image and the radio image would be expected
for excitation in cooling tails of SNRs.

\section{CONCLUSION}

ESO248-G10 has been modelled as a triaxial ellipsoid with axes ratios of
$B/A=0.75$ and $C/A=0.69$, oriented with angles
$\psi=34^{\circ}$, $\theta=65^{\circ}$, and $\phi=19^{\circ}$. This is not
an exclusive solution but one that is consistent with the observed apparent
axes and rotation angles.
The core has been identified as an AGN, surrounded by either hot dust emission or
dust extinction. Hot gas has been found to play an increasing role beyond a
radius of 4\,arcsec (6.5\,kpc).
An optical spectrum has revealed fast gas rotation out to a radius of 
20\,arcsec ($\sim33$\,kpc).
A slower velocity region extending from a radius of 5 to 15 arcsec to the north-east
has been identified as a second velocity feature. A model is proposed in which this
feature is a merging gas-rich galaxy, which is inducing star formation as it settles
into an orbit about the figure rotation axis.

\section*{ACKNOWLEDGMENTS}
We would like to thank Oak-Kyoung Park for taking the {\it J} and {\it H}$-$band
CASPIR images for us, and Elaine Sadler for helpful discussions and comments
on an earlier draft of the paper. This research has been supported in part
by a grant from the Australian Research Council.

\section{BIBLIOGRAPHY}

\noindent Allen, D. 1992, Proc. Astron. Soc. Aust., 10, 94\\
Alonso$-$Herrero, A., Simpson, C., Ward, M. J. \& Wilson, A. S. 1998, ApJ, 495, 196\\
Alonso$-$Herrero, A., Ward, M. J. \& Kotilainen, J. K. 1996, MNRAS, 278, 902\\
Arnaboldi, M., Capaccioli, M., Cappellaro, E., Held, E. V. \& Koribalski, B. 1995,
AJ, 110, 199\\
Barvainis, R. 1987, ApJ, 320,537\\
Bertola, F., Buson, L. M. \& Zeilinger, W. W. 1992, ApJ, 401, L79\\
Binney, J. 1978, MNRAS, 183, 779\\
Bryant, J. J. \& Hunstead, R. W. 1999, MNRAS, 308, 431\\
Capaccioli, M., Cappellaro, E., Held, E. V. \& Vietri M. 1993 A\&A, 274, 69\\
Capaccioli, M. \& Longo, G. 1994, ARA\&A, 5, 293\\
Carballo, R., Sanchez, S. F., Gonzalez$-$Serrano, J. I., Benn, C. R. \& Vigotti, M.
1998, AJ, 115, 1234 \\
de Zeeuw, P. T. \& Franx, M. 1989, ApJ, 343, 617\\
Fanaroff, B. L. \& Riley, J. M. 1974, MNRAS, 167, 31\\
Fisher, D., Illingworth, G. \& Franx, M. 1994, AJ, 107, 160\\
Gooch, R. 1996, ADASS, 5, 80\\
Hawarden, T. G., Elson, R. A. W., Longmore, A. J., Tritton, S. B. \& Corwin Jr., H. G.
1981, MNRAS, 196, 747\\
Heckman, T. M., et al. 1986, ApJ, 311, 526\\
Hill, T. L., Heisler, C. A., Sutherland, R. \& Hunstead, R. W. 1999, AJ, 117, 111\\
Hui, X., Ford, H. C., Freeman, K. C. \& Dopita, M. A. 1995, ApJ, 449, 592\\
Jones, P. A. 1989, PASA, 8, 81\\
Kormendy, J. \& Djorgovski, S. 1989, ARA\&A, 27, 235\\
Kotanyi, C. G. \& Ekers, R. D. 1979, A\&A, 73, L1\\
Kotilainen, J. K., Ward, M. J., Boisson, C., DePoy, D. L., Bryant, L. R. \& Smith, 
M. G. 1992, MNRAS, 256, 149\\
Merritt, D. \& de Zeeuw, T. 1983, ApJ, 267, L19\\
McGregor, P., Hart, J., Dowing, M., Hoadley, D. \& Bloxham, G. 1994, ExA, 3, 139\\
Mollenhoff, C.,Hummel, E. \& Bender, R. 1992, A\&A, 255, 35\\
Nicholson, R. A., Bland$-$Hawthorn, J. \& Taylor, K. 1992, ApJ, 387, 503\\
Radovich, M. \& Rafanelli, P. 1996, A\&A, 306, 97\\
Saripalli, L., Subrahmanyan, R. \& Hunstead, R. 1994, MNRAS, 269, 37\\
Sault, R. J., Teuben, P. J. \& Wright, M. C. H. 1995, ADASS, 4, 433\\
Shortridge, K. 1993, in ASP Conf. Ser. 52, Astronomical Data Analysis Software 
and Systems II, ed. R. J. Hanisch, R. J. V. Brissenden \& J. Barnes
(San Francisco: ASP), 219\\
Stark, A. A. 1977, ApJ, 213, 268\\
Steiman-Cameron, T. Y. \& Durisen, R. H. 1982, ApJ, 263 L51\\
Steiman-Cameron, T. Y., Kormendy, J. \& Durisen, R. H. 1992, AJ, 104, 1339\\
Surace, J. A. \& Sanders, D. B. 1999, ApJ, 512, 174\\
Tody, D. 1993, in ASP Conf. Ser. 52, Astronomical Data Analysis
Software and Systems II, eds. R. J. Hanisch,
R. J. V. Brissenden, \& J. Barnes
(San Francisco: ASP), 173\\
van Albada, T. S., Kotanyi,C. G. \& Schwarzschild, M. 1982, MNRAS, 198, 303\\

\end{document}